\documentclass[sigconf]{acmart}
\input{glyphtounicode}
\usepackage[utf8]{inputenc}

\setcopyright{none}
\renewcommand\footnotetextcopyrightpermission[1]{} 

\begin{document}

\title{The CASE Framework - A New Architecture for Participatory Research and Digital Health Surveillance}
\thanks{This version is a preprint uploaded to arXiv and has not been peer reviewed.}

\author{Marco Hirsch}
\affiliation{%
  \institution{German Research Center for Artificial Intelligence (DFKI)}
  \city{Kaiserslautern}
  \country{Germany}}
\email{marco.hirsch@dfki.de}

\author{Peter Hevesi}
\affiliation{%
  \institution{coneno GmbH}
  \city{Kaiserslautern}
  \country{Germany}}
\email{hevesi@coneno.com}

\author{Paul Lukowicz}
\affiliation{
\institution{University of Kaiserslautern-Landau (RPTU)}
  \institution{German Research Center for Artificial Intelligence (DFKI)}
 \city{Kaiserslautern}
  \country{Germany}}
\email{paul.lukowicz@dfki.de}

\begin{abstract}
We present CASE, an open-source framework for adaptive participatory research and disease surveillance. Unlike traditional survey platforms with static branching logic, CASE uses an event-driven architecture that adjusts survey workflows in real time based on participant responses, external data, temporal conditions, and evolving participant state. This design supports everything from simple one-time questionnaires to complex longitudinal studies with sophisticated conditional logic.

Built on over a decade of practical experience, CASE underwent major architectural changes in 2024. We replaced a complex microservice design with a streamlined monolithic architecture, significantly improving maintainability and deployment accessibility, particularly for institutions with limited technical resources.

CASE has been successfully deployed across diverse domains, powering national disease surveillance platforms, supporting post-COVID cohort studies, and enabling real-time sentiment analysis during political events. These applications, involving tens of thousands of participants, demonstrate the framework's scalability, versatility, and practical value.

This paper describes the foundations of CASE, documents its architectural evolution, and shares lessons learned from real-world deployments across diverse research domains and regulatory environments. We position CASE as a mature research infrastructure that balances sophisticated functionality with practical deployment needs for sustainable and institutionally controlled data collection systems.
\end{abstract}


 
\begin{teaserfigure}
  \includegraphics[width=\textwidth]{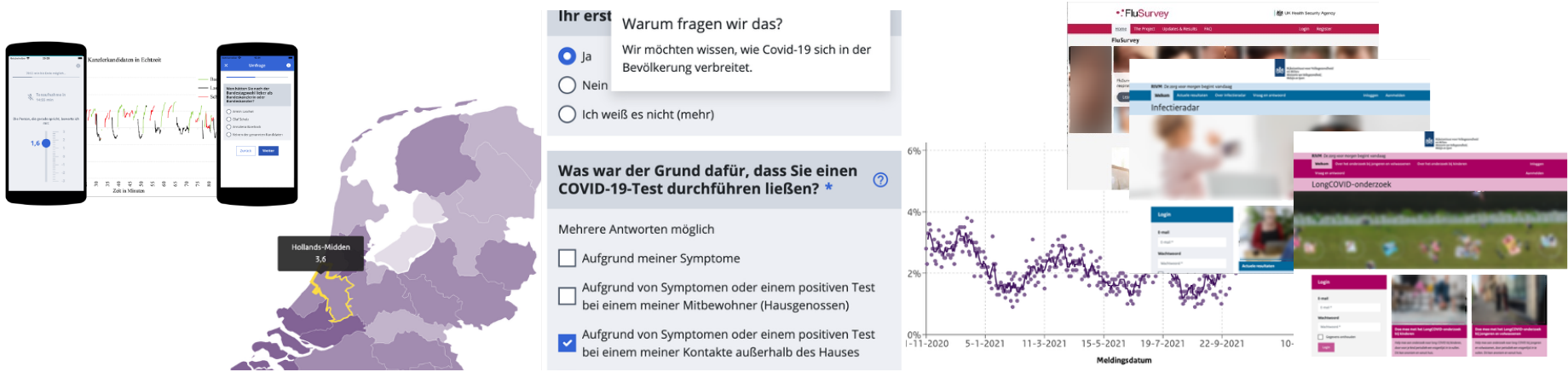}
  \caption{Screenshots from real-world applications built with the CASE framework.}
  \Description{Collage of screenshots from real-world CASE framework applications showing interfaces for surveys, participant dashboards, and study management tools.}
  \label{fig:teaser}
\end{teaserfigure}

\bibliographystyle{ACM-Reference-Format}

\maketitle 

\pagestyle{plain} 

\section{Introduction}

Digital survey platforms have become essential tools in participatory research with various different tools available across different domains~\cite{wright2017}. Many provide survey logic capabilities, typically allowing researchers to define conditional branching or skipping questions based on respondent answers. Many platforms also offer dynamic content features where responses can be piped into later questions, allowing basic personalization of the questionnaire experience. Despite this, most mainstream survey platforms still follow a mostly linear, predefined flow. The branching logic is usually limited to conditions based on previous responses within the survey itself. Truly context-aware or externally adaptive workflows remain rare. Recent analyses highlight that popular survey services "lack usable mechanisms for seamlessly importing participants' data from other systems"~\cite{velykoivanenko2024surveysystem}. Although several platforms can preload known information or trigger minor follow-ups, there is limited support for dynamically incorporating external context, such as real-time sensor data, environmental conditions, or personal digital data into the survey logic during execution.

CASE addresses these limitations through an event-driven architecture where surveys respond dynamically to multiple inputs rather than following predetermined paths. Unlike conventional survey tools that rely on static branching logic, CASE implements an event-driven architecture that enables dynamic and context-aware workflows. Rather than treating surveys as fixed sequences of questions, CASE models them as responsive systems that can adjust in real time based on multiple inputs, such as participant responses, temporal conditions (even ensuring minimal time distances to validate triggers), external data, or changes in user state. CASE tracks each participant's state continuously. The same triggers that advance a survey, like a submitted response, a timer event, or external data, can also update participant state. Survey flows can therefore diverge or iterate based on real-time conditions, making CASE suitable for longitudinal studies where context matters.

The practical value of CASE has been demonstrated through deployment in many different real-world scenarios. This shows that beyond its conceptual flexibility, CASE is grounded in real-world design needs, shaped by practical demands such as privacy, security, scalability, and long-term maintainability. During development, we prioritized not only advanced functionality, but also sustainability in deployment. The framework is designed to be reliably operated across diverse institutional and infrastructural settings.

The decision to undertake a comprehensive rework of the CASE framework in 2024 was driven by several critical factors that emerged from years of practical deployment experience. The growing number of platforms that utilize CASE technology revealed the need for greater flexibility to support various technical, ethical, legal, and use-case-specific requirements. Additionally, the goal of remaining aligned with modern web development standards and evolving research methodologies leads to a fundamental architectural reassessment. The rework replaced the previous complex microservice setup with a streamlined and simplified architecture, significantly reducing necessary code management efforts, reducing interdependencies, and improving ease of deployment, which is critical for institutions with limited technical capacity or specific infrastructure requirements. This further positions CASE as a modern, maintainable solution that balances sophisticated functionality with practical deployment options.

This paper makes three key contributions. First, we present the design and implementation of an event-driven architecture tailored for adaptive participatory research, enabling context-sensitive real-time survey workflows. 
Second, we document the architectural evolution of the CASE framework, from an initial microservice-based design to a simplified, maintainable monolithic architecture based on insights gained through real-world deployments.  This includes development of graphical configuration tools that enable study creation without programming expertise. 
Third, we share lessons learned from five years of real-world deployments across diverse research domains and regulatory environments, including national surveillance platforms, specialized health monitoring systems, and cross-domain applications, providing practical insight for future participatory research infrastructure. These deployments, which involve tens of thousands of participants, also serve to validate the scalability and adaptability of the framework.

The remainder of this paper is organized as follows. Section~\ref{section:background} outlines the foundation and history of the framework and discusses related work. Section~\ref{section:fwoverview} provides an overview of the key modules and features that make up the CASE framework. These are the result of extensive collaboration with domain experts over several years, with input from a variety of stakeholders influencing the evolution of the framework.  Section~\ref{section:architecture} details the architectural improvements that started in 2024, explaining the motivations behind the rework and the resulting technical choices and adaptations. Section~\ref{section:applications} presents selected real-world deployments and lessons learned from these implementations, providing practical insights for future participatory research infrastructure.

\section{Foundations and Related Work} \label{section:background}
\subsection{Background}

The theoretical foundations of the framework date back to 2017, building on an idea that proposed a fundamentally different approach to survey logic~\cite{damron2017}. This work introduced the concept of dynamic question selection through mathematical relevance, where questions would be chosen from a pool based on calculated relevance rather than predetermined sequences.
The original vision emphasized a generalized view of data that could incorporate not only participant responses, but also sensor readings, location context, and external data sources. Virtual sensors were proposed to enable the integration of machine learning algorithms and complex computational logic, laying the foundation for the developed event-driven architecture. Although the actual CASE architecture evolved from these initial concepts into a more complex system than originally envisioned, the fundamental principle remained: Moving beyond static linear questionnaires to create an adaptive research instrument. 

The first implementation of those concepts began back in 2018 and was driven by a practical need to update the aging \emph{Influenzanet} platform. This participatory surveillance system for influenza-like illnesses (ILI) has been in operation since 2003. Over the years, it has expanded into a network of national platforms in 10 European countries, collectively engaging tens of thousands of volunteers each season, providing valuable data for epidemiological research and disease monitoring~\cite{Koppeschaar2017, paolotti2014}. However, by the mid-2010s, the underlying software of the platform was becoming outdated and increasingly difficult to maintain. It was not designed to incorporate new diseases or modern data practices, such as smartphone input or flexible consent models or enhanced privacy protection. CASE took the role as the successor to this legacy software as a next generation framework to carry forward the citizen science concept of \emph{Influenzanet} on a more robust, scalable, and sustainable technology.

\subsection{Related Work}

The landscape of digital data collection and participatory research platforms has evolved significantly in the last two decades, with various systems that address different aspects of survey deployment, participant participation, and real-time aspects. This section examines existing general approaches or solutions within related domains and positions CASE within this broader ecosystem.

\paragraph{Traditional Digital Survey Platforms}

Most mainstream survey platforms like \emph{LimeSurvey}, \emph{SurveyMonkey}, or \emph{Qualtrics} provide a web-based questionnaire design with conditional branching and basic logic elements. \emph{LimeSurvey} offers an expression manager for defining boolean conditions~\cite{limesurvey2025expression}, while \emph{Qualtrics} supports complex survey flows with visibility logic and embedded data integration~\cite{qualtrics2025logic}. These tools improve classic paper surveys by allowing skip patterns and piping of answers into subsequent questions. However, they remain fundamentally linear with fixed tree structures, where branches depend only on previous responses. Integration of real-time external context is extremely limited and is typically restricted to preloading known data. Dynamic adaptation during ongoing surveys (e.g., altering question flow based on external input or evolving conditions) is not supported, especially when it involves participant context or state. Although relatively user-friendly and sufficient for basic research, they lack the event-driven architecture and live context integration needed for highly adaptive or longitudinal study designs. More sophisticated platforms like \emph{Alchemer} support API integration and custom scripting. However, they still rely on predefined survey structures and do not support a full reconfiguration of the survey logic in response to live context or changes in participant state~\cite{alchemer2025}.
Basic tools like \emph{Google Forms} and \emph{Microsoft Forms} offer simple survey creation, but lack advanced logic capabilities, and are primarily designed for casual data collection rather than research applications. These limitations become more apparent in mobile and field contexts, where researchers need offline capability and regulatory compliance.

\paragraph{Mobile and Field Data Collection Tools}
The ubiquity of smartphones has enabled an increase in platforms optimized for mobile and offline data collection, such as \emph{Open Data Kit} (ODK)~\cite{Hartung2010odk}, \emph{KoBoToolbox}, \emph{SurveyCTO}, or the widely used \emph{REDCap}~\cite{harris2009redcap}. These tools introduced GPS transmission, photo upload, and offline synchronization capabilities. ODK supports complex form logic that includes skip patterns, input validation, and calculated fields that enable conditional question displays within a form. \emph{REDCap} additionally offers regulatory compliance and longitudinal data collection modules for multi-visit clinical studies. Despite their strengths in reliable data capture and compliance, these platforms focus on static surveys. Adaptivity is limited to logic within forms. Once deployed, external data or events cannot alter the flow of the questionnaire. Implementing long-running or interactive studies requires manually scheduling separate surveys or custom workflows outside the core system. These platforms focus on robust, predefined form execution rather than adaptive, context-sensitive survey workflows that leverage evolving participant state and runtime conditions.

\paragraph{Context Aware and Sensor-Driven Surveys}
Researchers have explored context-aware survey frameworks that react to sensor data or environmental events. \emph{Intille} proposed early concepts of context-aware experience sampling~\cite{Intille2003}, while the \emph{MyExperience} framework provided systems to trigger in situ questionnaires based on sensor readings~\cite{Froehlich2007}. Studies demonstrated that mobile sensors could initiate relevant questions at fitting moments, for example, asking a user when the accelerometer detects physical activity or when the GPS indicates arrival at a location, improving data relevance and accuracy~\cite{Srinivas2019}. In the initial phase of CASE development, the use of smartphone sensors was explored to enhance participatory research capabilities~\cite{hirsch2018grippenet}, capturing movement patterns and environmental context to augment traditional self-reported responses. However, such systems, while widely explored~\cite{Cornet2018}, remained largely proof-of-concept rather than deployed platforms. Implementation required significant custom programming and tight coupling of survey logic with mobile apps. Technical challenges increased as mobile operating systems introduced stricter privacy controls and background processing restrictions~\cite{petter2019crowdsensing}. Earlier context-aware systems focused on momentary interactions rather than providing infrastructure for multi-year or large cohort studies. Recent research continues to seek adaptive survey solutions, including data-driven survey generation approaches~\cite{velykoivanenko2024surveysystem}, highlighting the ongoing demand for platforms that intelligently adapt to the context of the user. However, a gap remains between research prototypes and robust general-purpose frameworks for longitudinal studies.

\paragraph{Mobile Health Research Platforms}
Dedicated mobile health research frameworks like \emph{Apple's ResearchKit}~\cite{researchkit2025}, \emph{Google's ResearchStack}~\cite{researchstack2025} represent specialized tools for longitudinal health studies. ResearchKit has enabled large-scale studies such as the \emph{Stanford Heart Study} and \emph{Parkinson's mPower} study, demonstrating the potential for mobile-based participatory research~\cite{Bot2016, McConnell2017}. These platforms provide built-in consent frameworks, sensor data integration, and standardized health survey modules. However, these frameworks are designed primarily for specific mobile ecosystems and health research contexts. They lack cross-platform deployment and are limited in their ability to handle diverse research domains beyond health. Furthermore, while they support some sensor integration, they do not provide the event-driven context-aware survey logic that enables real-time adaptation based on external data sources or complex management of the state of participants in multiple studies~\cite{Pratap2020}.

\paragraph{Participatory Surveillance Platforms}
Several participatory surveillance platforms have demonstrated the value of crowd-sourced symptom reporting for the monitoring of infectious diseases. In the US, \emph{Flu Near You} and its successor \emph{Outbreaks Near Me}~\cite{outbreaknearme2025} allowed tens of thousands of volunteers to report influenza-like and COVID-19 symptoms weekly, although their static technical architecture limits dynamic survey adaptation, remains largely undocumented and was not designed for broader reuse~\cite{Smolinski2015}. Australia's \emph{FluTracking}~\cite{Carlson2019} has similarly shown high public engagement and epidemiological value during seasonal influenza outbreaks, but, like other single-purpose platforms, lacks the technical flexibility to dynamically adapt survey flows or rapidly adapt to new diseases during emerging health threats. The UK's \emph{ZOE COVID Symptom Study} demonstrated the scalability and research potential of participatory symptom tracking at unprecedented scale, engaging millions of users and generating findings that directly influenced public health policy~\cite{Menni2020}. Beyond its initial role in tracking COVID-19 symptoms, the ZOE app has evolved into a broader research platform now known as the ZOE Health Study. It engages participants in longitudinal studies of diet, chronic symptoms, and post-COVID-19 syndrome~\cite{Rjoob2025,Bermingham2024}. This illustrates the potential for participatory research platforms to remain relevant beyond acute outbreaks. CASE supports a reusable and adaptive architecture that facilitates such transitions, enabling research continuity across domains. Many European platforms within the \emph{Influenzanet} network have since transitioned to the open-source CASE framework, establishing a reusable foundation for multi-study participatory research. This shift reflects a broader trend toward scalable, transparent, and adaptable infrastructures for participatory surveillance, identified as a critical need across global systems~\cite{McNeil2022}.

\paragraph{Summary}
The review of existing platforms reveals several key gaps that CASE addresses. Traditional survey platforms like \emph{LimeSurvey} and \emph{Qualtrics}, while user-friendly, lack real-time adaptiveness and external context integration. Mobile data collection tools such as \emph{ODK} and \emph{REDCap} focus on form-based data capture but are not able to dynamically adjust to external events once deployed. Context-aware research prototypes demonstrate promising concepts, but remain largely a proof-of-concept. Mobile health research platforms such as \emph{ResearchKit} are sophisticated, but limited to specific ecosystems and health domains. Furthermore, many commercial platforms have restrictive pricing models and limited customization options for research use cases~\cite{wright2017, sue2012conducting}.
CASE contributes to this landscape by implementing an event-driven architecture that enables dynamic survey workflows based on external context, participant state, and temporal factors. The framework has been deployed in several participatory surveillance systems and longitudinal studies. Although primarily applied in health research contexts, its modular design allows adaptation to other domains, as illustrated by its use in analysis of political sentiment during live events. This flexibility positions CASE not only as a platform for health-related studies, but also as a foundation for participatory research more generally. However, like other research frameworks, CASE still requires technical expertise for deployment and customization to specific research needs.

\section{Framework Overview and Requirements}\label{section:fwoverview}
\subsection{Requirements Context and Scope}

The features outlined in this section were derived from a comprehensive set of requirements gathered through interviews and discussions with the application owners, who are recognized experts in their respective fields of participatory studies. Structured requirement engineering~\cite{requirmentsengineering} is essential for successful system development. However, the underlying rationale for these requirements is outside the scope of this work, and we focus on presenting them as the foundational parameters that guided our development process.  This paper shows how we implemented solutions to effectively meet the needs set by experts, highlighting the technical realization of fulfilling the requirements.

Through our experience in implementing diverse applications, collaborating with experts, and deploying studies across various contexts, we have made several key observations. Although many applications share common components and logic, each typically comes with a unique set of requirements that can occasionally conflict. In addition, applications deployed in different countries often need to comply with varying regulatory requirements. Another observation is that user interface designs must adopt different stylistic approaches and content goals depending on the specific goals, contexts, and target audiences involved.

In response to these observations, our primary goal has been to develop a technical framework that allows for the rapid composition of tailored applications. This framework aims to maximize the reusability of common logic and components while simultaneously providing the flexibility to accommodate the specific set of requirements of individual use cases.

The features described in the following subsections reflect our effort to balance standardization and customization, creating a system that is both efficient and adaptable to diverse needs in participatory studies.

\subsection{Overall Application Goal}
The framework was designed to support a wide range of participatory studies, from simple one-time surveys to more complex long-term research projects. Its flexible architecture addresses the needs of both study participants and researchers. The main objectives of the CASE framework can be summarized as follows:

\paragraph{Ethical Design}
The whole system was designed with explicit attention to the principles of data minimization and transparency. CASE supports state-of-the-art encryption, role-based data access and is capable of implementing consent workflows informed by established ethical guidelines for digital participatory surveillance~\cite{Genevieve2019ethics, who2022ethics} which include the need for clear, multilingual and user-friendly electronic consent procedures to ensure participant autonomy and data protection.

\paragraph{Flexibility in Study Design}
The framework is capable of supporting diverse types of studies. On the one hand, it can manage basic anonymous surveys that require minimal engagement. However, it can handle complex longitudinal studies, in which participants are actively managed over extended periods, often participating in multiple studies with continuous follow-up. The flexibility of the framework allows it to be tailored to different research fields and study complexities, making it suitable for use in various domains.

\paragraph{Dual-Focus Functionality}

The framework is built around two primary components. The \emph{participant interface} provides a user-friendly way for individuals to complete surveys, track participation, and interact with researchers. The design prioritizes simplicity and engagement, ensuring that participants can easily navigate the interface and maintain long-term participation in the studies. 
Meanwhile, \emph{researcher and manager interface} offers a robust system that allows researchers and application managers to design, execute, and monitor studies. It includes tools for study configuration, survey customization, participation tracking, and data analysis, ensuring researchers' control to manage the full study lifecycle. 
By focusing on both participant experience and researcher management needs, the CASE framework bridges the gap between efficient data collection and study administration.

\subsection{Core Functional Modules}
The framework consists of several key modules, each designed to handle a specific aspect of participatory studies. These modules collectively provide a comprehensive and flexible solution that can be customized to meet the unique needs of various research endeavors.

\subsubsection{Study System}

The module \emph{study system} is a key component of the framework, offering a robust and highly configurable environment to manage complex studies. It is based on a sophisticated event-driven engine that serves as the core component to manage the flow and logic of the study. The key features of the system include the following.

\begin{itemize}
    \item Event-Driven Execution: The module operates on a set of configurable rules that determine how the system reacts to different events, like participant enrollment, response submission, or periodic timers checking states or custom events triggered by the specific application.
    \item Context-Based Decision Making: The system can utilize various context sources, such as certain participant responses, response history, current participant state, or payload data from external events.
    \item Customizable Action Logic: Study managers can define rules that trigger actions in response to events, such as assigning new surveys, scheduling messages, or sending data to external systems or event handlers. 
\end{itemize}

\begin{figure}[ht]
  \includegraphics[width=\columnwidth]{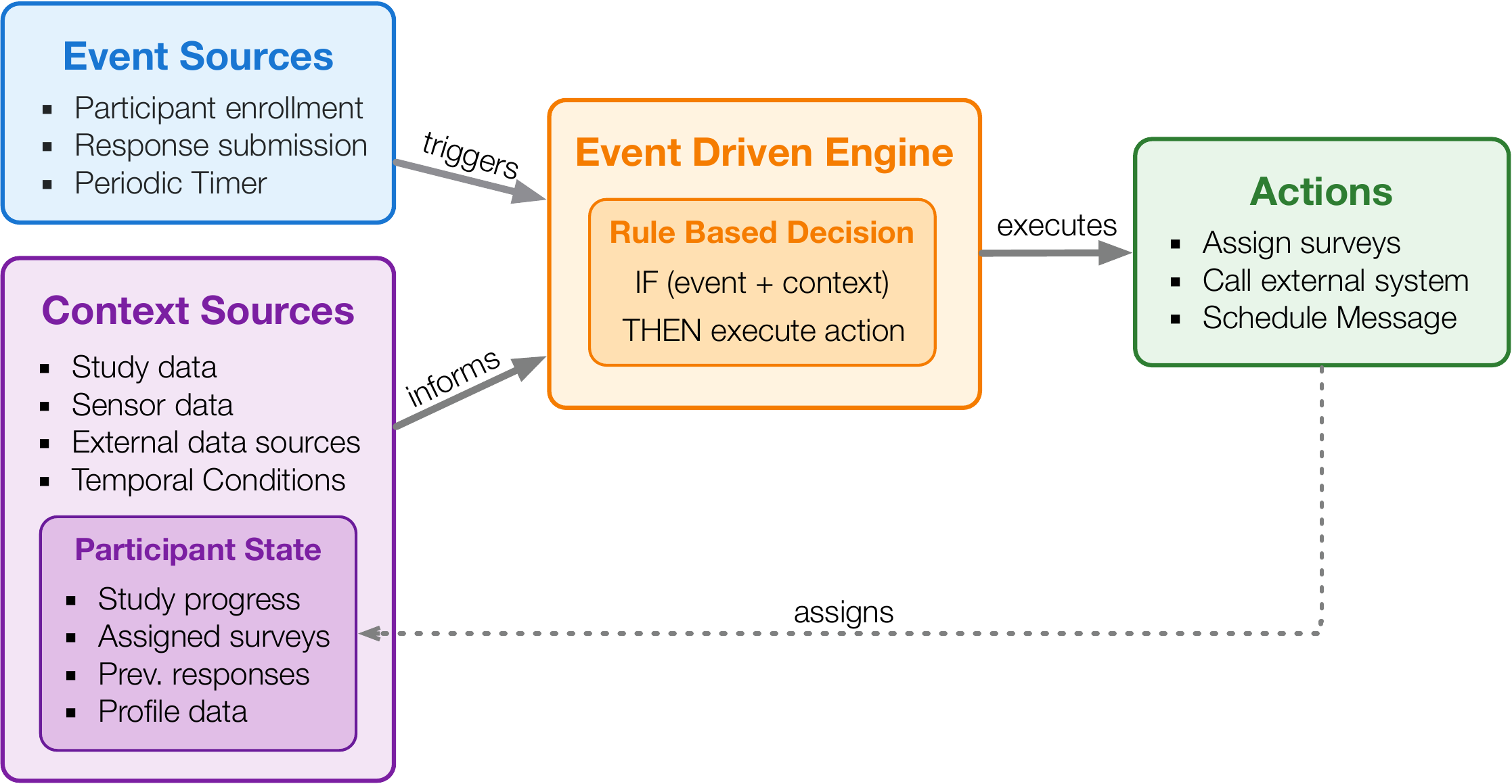}
  \caption{A diagram showing the architecture of the CASE framework's \emph{study system}. Participant events, timers, context sources (such as sensor data or external inputs), and the participant's state (including response history) are processed by a rule-based engine to trigger actions like survey assignments, message scheduling, or external system calls.}
  \Description{The diagram is a flowchart illustrating the event-driven architecture of the CASE study system. On the left, event sources such as participant enrollment, response submission, and periodic timers are shown. Below, context sources include study data, sensor data, external data sources, and temporal conditions. A central rule-based engine processes these inputs along with participant state information, such as assigned surveys, previous responses, and profile data, shown on the right. Arrows indicate that these inputs inform decisions that result in actions like assigning surveys, scheduling messages, or calling external systems.}
  \label{fig:study-system}
\end{figure}

Figure~\ref{fig:study-system} illustrates the underlying architecture of the \emph{study system}, showing how events, context, and participant state flow into a rule-based engine that drives these actions. This flexibility allows for the management of complex and dynamic studies with individualized participant interactions. It is particularly well suited for large-scale or long-term studies in fields such as epidemiology or public health, where ongoing engagement, adaptability, and automation of certain processes are crucial.

\subsubsection{Survey Module}

The \emph{survey module} supports the creation of context-aware, dynamic surveys that can adapt in real time based on participant's responses, improving the quality and relevance of collected data. Key features of the survey module include:

\begin{itemize}
    \item Survey Renderer: A flexible renderer that supports a variety of commonly used question types and other items, like formatted text, ensuring that surveys are both visually appealing and functional.
    \item Survey Engine: Responsible for resolving dynamic expressions defined by researchers, enabling adaptive questionnaire behavior based on real-time evaluation of predefined logic.
    \item Expression System: Expressions can incorporate context variables from multiple flexible sources, including previous participant responses, sensor data, external data sources, or current response states, providing comprehensive adaptability to participant context and environmental factors.
    \item Dynamic Content Generation: Resolved expressions enable dynamic content within surveys, such as generating relative dates, displaying response counts, or incorporating calculated values that update based on participant interactions or external conditions.
    \item Conditional Logic Control: Expression resolution drives sophisticated conditional logic that can dynamically display or hide individual survey items or only their components, groups of items, as well as enable or disable interactive components based on evaluated conditions.
\end{itemize}

\begin{figure}[ht]
  \includegraphics[width=\columnwidth]{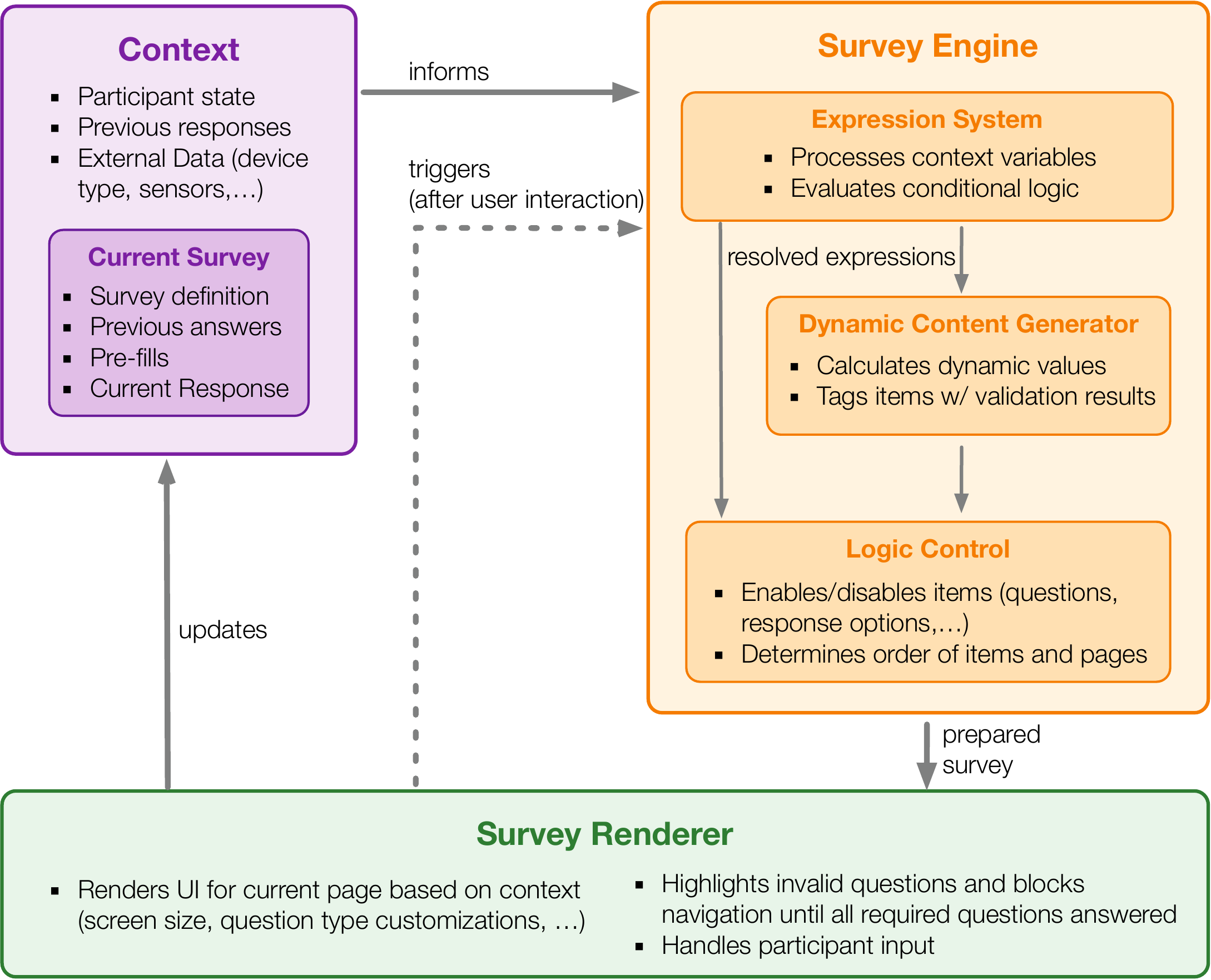}
  \caption{A diagram showing the architecture of the CASE framework's \emph{survey module}. Context sources including survey definitions, previous and current responses inform the Survey Engine, which processes expressions, generates dynamic content, and controls survey logic to produce an adaptive, context-aware survey experience via the \emph{survey renderer}.}
  \Description{The diagram illustrates the Survey Module's architecture with three main components. On the left, the Context box contains participant state, previous responses, external data (device, sensors), and current survey information including survey definition, previous answers, pre-fills, and current response. In the center, the Survey Engine comprises three sub-components: the Expression Solver (which processes context variables and evaluates conditional logic), the Dynamic Content Generator (which calculates dynamic values and tags items with validation results), and the Logic Control (which enables/disables items and determines order of items and pages). These components process information sequentially, with resolved expressions flowing from the Expression Solver through to Logic Control. On the right, the Survey Renderer renders the UI for the current page based on context, highlights invalid questions, blocks navigation until all required questions are answered, and handles participant input. Arrows show information flow: context informs the Survey Engine, which outputs a configured survey to the Renderer, and user interactions trigger updates back to the context, creating a feedback loop.}
  \label{fig:survey-module}
\end{figure}

Figure~\ref{fig:survey-module} illustrates the architecture of the \emph{survey module}, detailing how data is processed and the listed components interact with each other. This adaptability makes the survey module an effective tool for maintaining a high level of participant engagement while collecting high-quality research data in complex environments via context-sensitive questionnaires.

\subsubsection{Authentication and User Management}

Authentication and user management are crucial for tracking participants across multiple surveys over time and ensuring secure access to the system. This module supports both the participant and researcher roles, ensuring appropriate access control and data protection. Key features include:

\begin{itemize}
    \item Participant Authentication: The built-in module primarily supports authentication through email and password. As an additional option, temporary one-time codes are supported to further protect sensitive resources.
    \item Participant Account Management: Participants have complete control over their accounts, allowing them to update email addresses, change passwords, and manage other account-related information.
    \item Researcher/Administration Authentication and Management: The system supports OpenID Connect for authentication to the management area. A permission system allows administrators to manage access rights for different researchers and management users. Restricted access can be granted for example to access response data or managing configurations and content, including study or survey management.
\end{itemize}

While this module was designed for composing more complex use cases, it is possible to build an application with the CASE framework incorporating other authentication solutions.

\subsubsection{Messaging System}

The framework provides comprehensive messaging functionality to support various use cases in research studies and surveys. These features enable efficient communication with participants. Key features include:

\begin{itemize}
    
    \item Template-Based Emails: The framework contains data models and logic to set up template-based emails that allow personalized messages with dynamic content. This includes insertions, such as authentication codes or other information based on participant data or study events.
    \item Message Scheduling: The system allows for scheduling various types of communication, such as newsletters, reminders, or participation invitations. Built-in mechanisms are in place to balance the load based on available system resources and avoid delivery issues like false spam detection through flood messaging.
\end{itemize}

The system can be extended to utilize other communication channels and paradigms. A built-in HTTP to SMTP bridge allows easy integration into existing email infrastructure.

\subsubsection{Configuration Tools} \label{section:config-tools}

Early versions of CASE required researchers to manually code survey definitions and study logic, creating barriers for teams without programming expertise. To address this, we developed graphical configuration tools that enable study creation and management through visual interfaces. These tools were developed as part of ongoing platform improvements and are publicly available as part of the CASE framework.

\paragraph{Survey Editor}
Survey creation previously required manual coding of survey definitions but can now be managed through a graphical interface with structured item editing and a built-in simulator. The Survey Editor allows researchers to construct surveys by arranging items hierarchically, like questions, display text, instructional content, and groups without writing code. 

Key features include:
\begin{itemize}
    \item Visual item arrangement with drag-and-drop reordering
    \item Multilingual content management for international studies  
    \item Built-in simulator for testing survey behavior before deployment
    \item Real-time preview showing how surveys will appear to participants
\end{itemize}

\paragraph{Expression Editor}
Rules and display conditions can now be created with a visual tool rather than manual scripting, speeding up iteration. The Expression Editor represents logical expressions as nested functions in a graphical interface. Researchers configure conditions, validation rules, and dynamic behavior by selecting functions and setting parameters visually.

For example, to show a follow-up question only when a participant selected "Other" in a previous dropdown, the researcher selects "Response contains any of these keys" and visually specifies which question and option to check. The tool generates the underlying expression automatically. This approach makes complex conditional logic accessible to researchers without programming background.

\begin{figure}[ht]
  \includegraphics[width=\columnwidth]{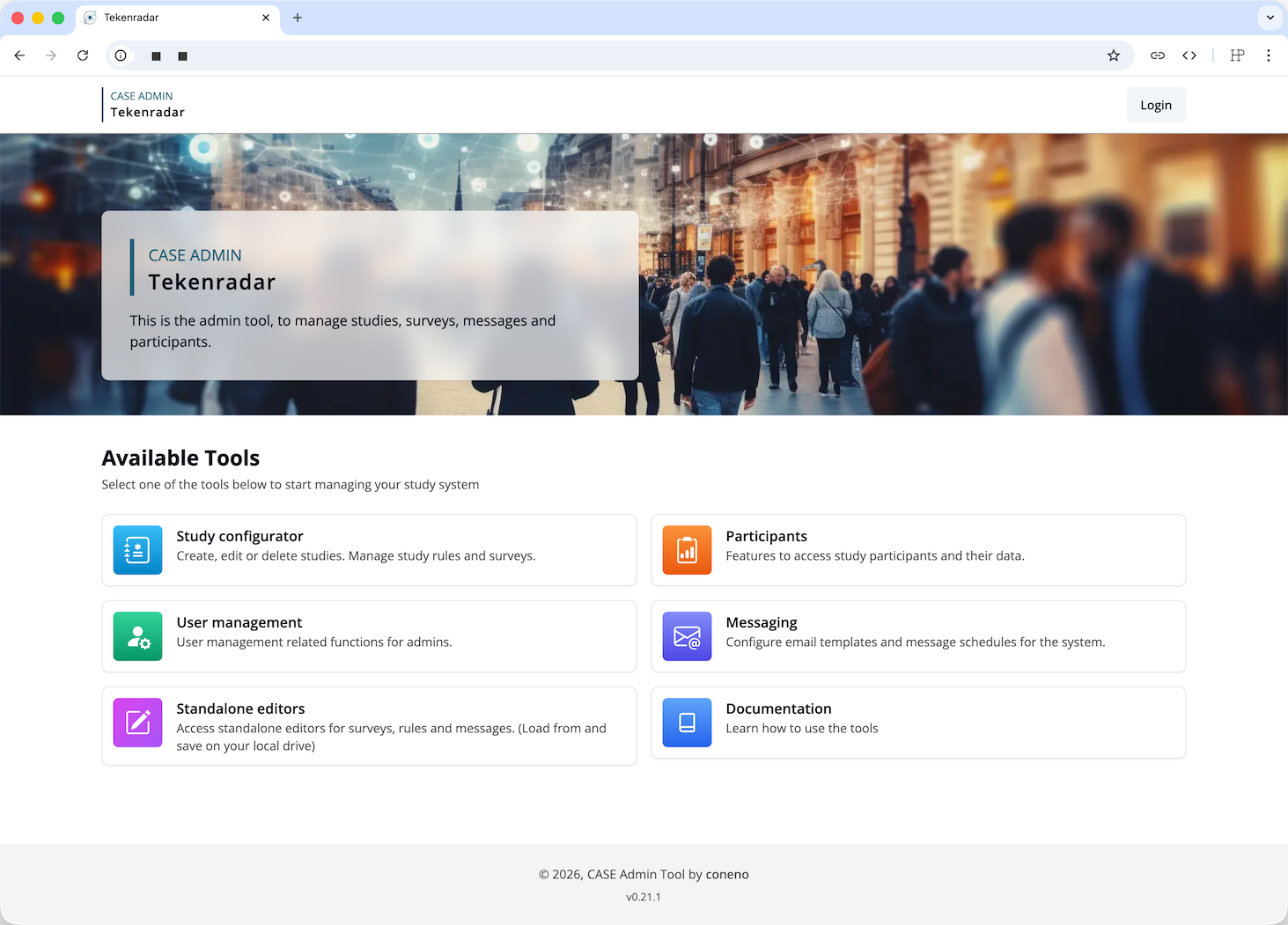}
  \caption{CASE Admin UI main menu, providing access to study configuration, participant management, messaging and editors for surveys and expressions. This interface addresses the need for non-technical teams to configure studies without manual coding.}
  \Description{Screenshot of the CASE Admin UI showing a modern web interface with a card-based layout. The main screen displays six tools: Study configurator, Participants, User management, Messaging, Standalone editors, and Documentation. Each is represented by an icon and brief description.}
  \label{fig:admin-ui}
\end{figure}

\paragraph{Admin UI}
A unified web interface consolidates study configuration, participant management, messaging templates, and access control. The Admin UI provides direct access to the Survey Editor, Expression Editor, and other configuration tools through a card-based dashboard (Figure~\ref{fig:admin-ui}). It integrates all study management functions into a single interface, reducing the need to interact with multiple systems or command-line tools.

While these tools significantly lower technical barriers for study configuration and management, deployment and initial platform setup still require institutional IT support. The tools focus on the operational phase, like creating studies, managing participants, configuring surveys rather than infrastructure deployment.

\section{Architecture and Implementation}\label{section:architecture}

\begin{figure}[ht]
  \includegraphics[width=\columnwidth]{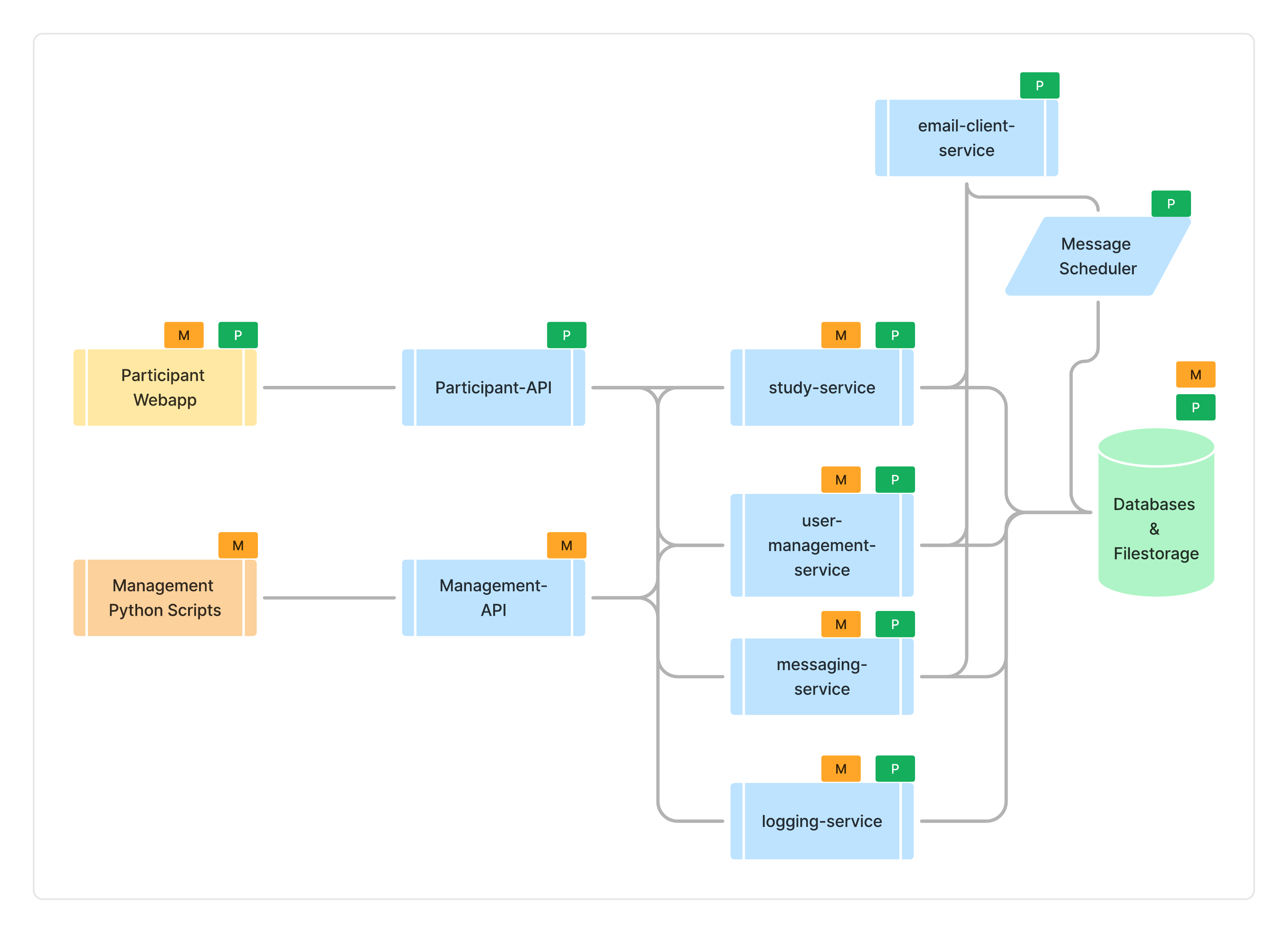}
  \caption{Thematic microservice architecture, with standalone modules organized by functional responsibility. Orange M labels components required for study management and configuration, while green P label modules necessary for participant facing functionality.}
  \Description{Diagram showing the previous CASE framework architecture using a microservice approach. It includes separate components for participant-facing features, study management, messaging, user management, and email services. The architecture involves multiple interdependent services, each handling specific functions, with APIs coordinating communication between them.}
  \label{fig:microservices}
\end{figure}

Between 2020 and early 2024, CASE-based applications were implemented using a microservice architecture. This approach aimed to provide modularity and scalability, reflecting best practices in cloud-native design at that time~\cite{Oyeniran2024}. An overview of the original architecture is shown in Figure~\ref{fig:microservices}. The applications were composed of the following main components:

\subsection{Previous Microservice Approach}

\paragraph{Front-End}
A web application implemented as a single-page React application, based on a shared application library. This library contains common data types, logic for data fetching, state management, navigation, and styling to support different use cases. 

\paragraph{Back-End}
The back-end was composed of the following core microservices implemented in the \emph{Go} programming language.
\begin{itemize}
  \item User management service
  \item Study service
  \item Messaging service
  \item Email client service
  \item Logging service
\end{itemize}

Each service is organized along thematic lines. Individual services were responsible for all methods related to specific functional areas of the system. Both management and participant-facing features were integrated within all relevant services, rather than being separated. These services exposed their functionality through gRPC interfaces, enabling structured communication between components.
In addition to the core services, two API services existed:
\begin{itemize}
  \item Management API service: Exposed back-end functionality for management operations
  \item Participant API service: Provided interfaces for participant-facing functions
\end{itemize}

Both API services acted as proxies, exposing the back-end functionality over HTTP webserver and routing incoming requests to the appropriate microservices.

\subsection{Disadvantages of the Microservice Approach}

While the initially anticipated benefits of microservice architecture for modularity and scalability did not materialize, as practical deployments evolved, several disadvantages became apparent. 
The code was distributed across multiple repositories, leading to unclear dependencies and relationships between modules, which made understanding the overall architecture of the system more challenging. 
Additionally, composing new applications from parts of the system proved more difficult, as the service parts focused more on their responsibility than offering composable modules.
Furthermore, simple changes often required modifications in multiple repositories, with corresponding API changes frequently necessary. 
The entanglement of management and participant functions within the services created additional complications, leading to the necessity for a full deployment of all services to begin the setup. This inability to deploy services independently reduced flexibility and increased deployment complexity.

These challenges emerged during sustained operation across multiple production deployments with different institutional partners, rather than from purely theoretical considerations. They are also not limited to CASE~\cite{Krug2024Emicroservice} but indicate the need for a reevaluation of system architectures based on the actual requirements and usage patterns of participatory research infrastructures.

\subsection{Simplified Architecture and Re-implementation}

In 2024, we performed a comprehensive architectural rework to address the previously mentioned issues. The new back-end implementation still uses the \emph{Go} language, as its properties make it highly suitable for the task. However, the new back-end combines all code, including data types, modules for feature areas, and database adapters, into a single monorepo. Figure~\ref{fig:simplified} shows an overview of this new architecture.  This approach significantly simplifies code management and intermodule dependencies. The new system includes reference implementations for the following key components:

\begin{figure}[ht]
  \includegraphics[width=\columnwidth]{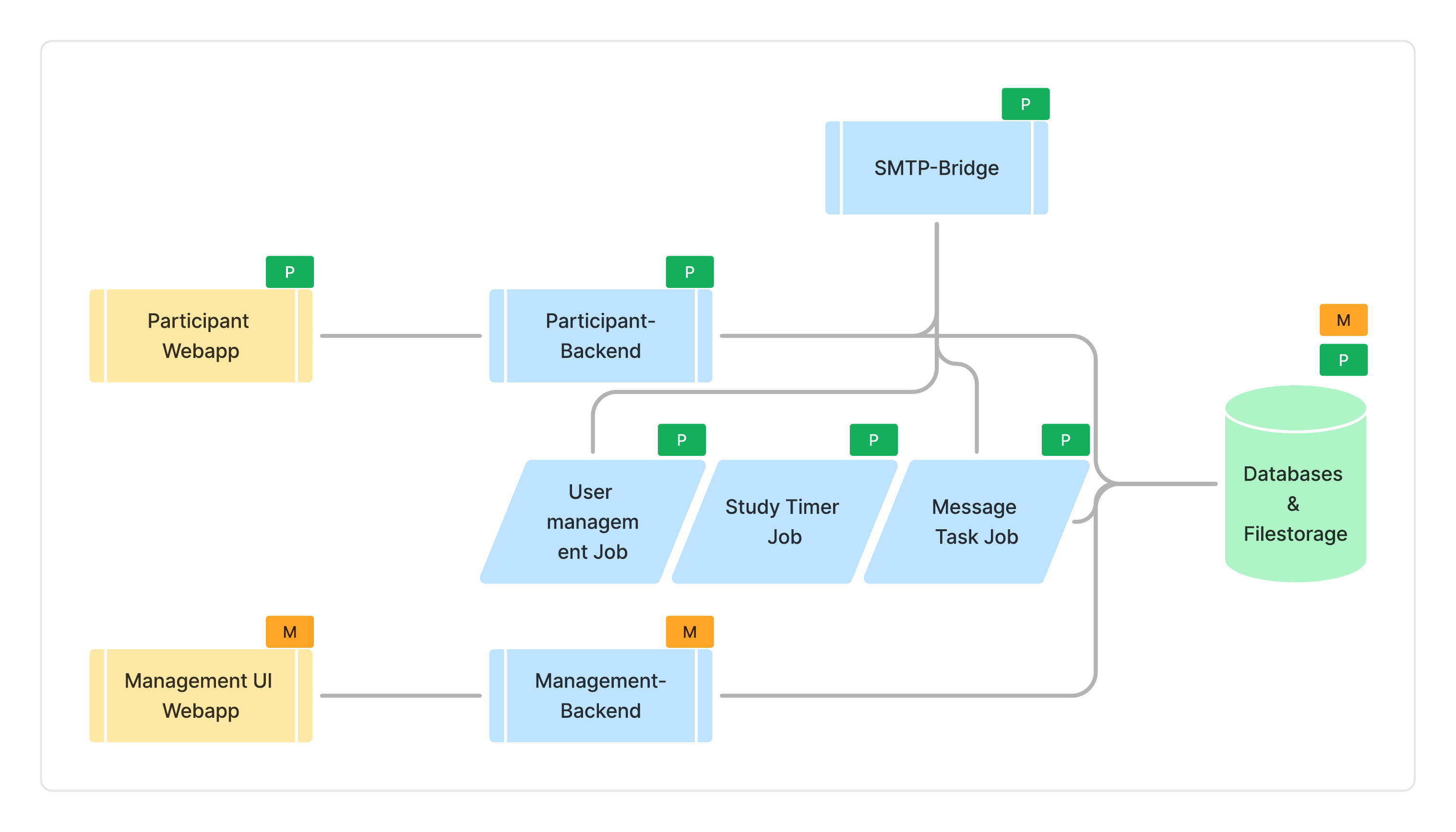}
  \caption{Simplified architecture optimized for easier deployment. Compared to Figure~\ref{fig:microservices}, it reduces interdependencies between components and lowers the complexity of setup. Green P labels mark modules required for participant facing applications  while orange M labels represent components needed for management functionality.}
  \Description{Diagram of the updated CASE framework architecture with simplified structure. It shows consolidated components for participants and researchers, such as a participant back-end, a management back-end, and schedulable jobs for user and study management. This new design reduces complexity by combining functions into fewer components.}
  \label{fig:simplified}
\end{figure}

\begin{itemize}
  \item Participant Back-End: A web server handling all participant facing features of the participatory study system. It encompasses authentication, study flows, and account management functionality.
  \item Management Back-End: A web server offering API functionality for administrative tasks. It allows for the configuration and management of study setups, message templates, and schedules. It also provides access to data and implements a resource-scope-based permission system.
  \item Schedulable Auxiliary Tasks: We have implemented several small programs that can be scheduled to run regularly on a fixed schedule. These include cleaning up unverified or inactive users, handling message tasks, and managing timer-based study events.
\end{itemize}

The reworked system also leverages modern front-end technologies, driven by the increased popularity of server-side rendering with React and the stable release of \emph{Next.js}'s app router approach. These advancements in front-end implementation brought several benefits, most notable for us:

\begin{itemize}
  \item Smaller bundle sizes
  \item Faster page loads
  \item Simpler state management
\end{itemize}

Additionally, new mature libraries emerged that facilitate the implementation of accessible user interfaces, such as \emph{Shadcn}/UI, based on \emph{Radix} UI and \emph{Tailwind} CSS.

This simplified architecture addresses many of the challenges that we faced with the previous microservice approach. By consolidating our code base and leveraging modern web technologies, we have created a more maintainable, flexible, and efficient system. The monorepo structure supports easier updates and better code reuse, while the separation of participant and management back-ends allows for more targeted development and deployment.

The new implementation uses backward compatible data models, and the new approach can be rolled out gradually. The microservice approach is still in use for multiple use cases and is currently not deprecated. The new approach is meant to provide an easier to deploy and maintainable version for use cases where this is relevant and suitable.

\section{Practical Information and Real-World Applications}\label{section:applications}

The CASE framework is openly available at \url{https://github.com/case-framework} where the source code and documentation can be accessed. It is licensed under the Apache 2.0 open-source license, ensuring transparency and adaptability for diverse environments. Development is primarily led by \emph{coneno GmbH}, with continued collaboration from academic and research partners.

In the following, we highlight notable research projects working with and on CASE, followed by selected key use cases, illustrating how the framework has been successfully implemented across diverse domains, from long-term participatory health studies to real-time data collection in mobile applications. We also include selected lessons learned, reflecting practical considerations observed during development and deployment.

\subsection{Selected Research Projects}\label{section:projects}

The CIMPLEX project (Bringing CItizens, Models and Data together in Participatory, Interactive SociaL EXploratories, 2014-2017, Horizon 2020, 9 partners)\footnote{CIMPLEX https://cordis.europa.eu/project/id/641191} developed the \emph{GrippeNet} App, a mobile application allowing self-reporting of symptoms to \emph{Influenzanet} enhanced with sensor-based features to analyze behavioral patterns during epidemics. CIMPLEX also initiated the transition from the older \emph{Influenzanet} technology stack to the newly developed, more advanced CASE framework.

Building on this foundation, the \emph{EpiPose} project (Epidemic intelligence to minimize 2019-nCoV's public health, economic and social impact in Europe, 2020-2023, Horizon 2020, 6 partners)\footnote{EpiPose  https://cordis.europa.eu/project/id/101003688} leveraged CASE technologies to develop new and improve existing platforms, allowing the collection and integration of COVID-19 related data from citizens. The project helped expand participatory surveillance systems such as \emph{Infectieradar} (in the Netherlands and Belgium) and \emph{Influweb} (in Italy), making them more efficient in monitoring a broader range of infectious diseases apart from influenza through citizen contributed health information. Additionally, \emph{EpiPose} contributed to the development of CASE customizations for the \emph{Influenzanet} infrastructure to fit specific national use cases.

The VERDI project (SARS-coV2 variants Evaluation in pRegnancy and paeDIatrics cohorts, since 2021-2025, Horizon Europe, 30+ partners)\footnote{VERDI https://cordis.europa.eu/project/id/101045989} uses CASE as a reference infrastructure to investigate how participatory surveillance systems can support pandemic preparedness under real-world constraints. VERDI is investigating the technological and methodological foundations required for resilient, scalable, and engaging data collection infrastructures that can quickly adapt to newly emerging infectious disease threats.

\subsection{Selected Use Cases}\label{sec:usecases}

For flu and COVID-19 surveillance, the \emph{infectieradar.nl} (Netherlands) and \emph{flusurvey.net} (UK) platforms, both part of the \emph{Influenzanet} network, demonstrate the capabilities of the CASE framework for large-scale participatory disease monitoring. These platforms engage tens of thousands of volunteers in weekly symptom reporting to track influenza, COVID-19, and other infectious diseases~\cite{McDonald2021,Koppeschaar2017}. In addition to seasonal surveillance, they also support ad hoc one-time surveys hosted within the same application ecosystem. 

Table~\ref{table:deployments} presents selected CASE deployments with verified participant statistics, demonstrating the framework's capacity to support sustained citizen engagement across different research contexts. Weekly participant numbers reflect 2024-2025 influenza season averages from Influenzanet.info data explorer~\cite{influenzanetdata}. Tekenradar reports weekly participants and yearly incident counts~\cite{NatureToday2025Tekenradar,RIVM2025TickRemoval}. The RTR participant count from 2021 and 2025 German federal election study combined~\cite{Maier2022,TVDuell2025}. Additional CASE-based platforms operate in Belgium, Switzerland, Italy, Spain, Estonia, and Portugal but current participation statistics were not publicly available at time of writing.

\begin{table}[th]
\centering
\caption{Selected CASE-based participatory surveillance deployments showing continuous operational scale.}
\label{table:deployments}
\begin{tabular}{llr}
\hline
\textbf{Platform} & \textbf{Country} & \textbf{Active Participants} \\
\hline
Infectieradar & NL & 13,700/week (peak: 37,700 in 2020)\\
Grippenet & FR & 3,400/week \\
Flusurvey & UK & 1,900/week \\
Tekenradar & NL & 350/week (6,000 tick bites in 2024) \\
RTR App & DE & 2,500 (debates 2021+2025 combined) \\
\hline
\end{tabular}
\end{table}

Beyond these documented deployments, additional European platforms within the \emph{Influenzanet} network have successfully transitioned from legacy technical infrastructure to CASE-based implementations, including 
\emph{infectieradar.be} (Belgium), 
\emph{gripiradar.ee} (Estonia), 
\emph{grippenet.fr} (France), 
\emph{influweb.org} (Italy), 
\emph{gripenet.pt} (Portugal), 
\emph{epidemiradar.es} (Spain), and 
\emph{grippenet.ch} (Switzerland).
While the open-source nature of CASE enables such migrations, not all implementations 
provide participation statistics or are publicly documented. These platforms operate independently with customized implementations of the CASE framework, tailored to their specific national contexts and requirements.

The Dutch platform \emph{tekenradar.nl} allows public participation in tracking tick bites and associated health effects for the monitoring of tick-borne diseases. The system has gathered tens of thousands of tick bite reports and supports epidemiological research on Lyme disease and related conditions through flexible survey and notification modules tailored to the experiences reported by individual participants~\cite{Garcia-Marti2018,Hofhuis2017}. Recently rebuilt with the CASE framework, it allows users to report tick encounters and symptoms while supporting follow-up engagement through a multitrack longitudinal study structure.

The Post-COVID Research Portal at \emph{postcovidonderzoek.nl} serves as a national entry point for individuals with long-term symptoms after COVID-19 infection to participate in ongoing scientific studies~\cite{RIVM2025PostcovidPortal}. Developed from the start with the reworked version of the CASE software, the platform collects baseline and follow-up data while enabling centralized participant recruitment across multiple coordinated research projects. By streamlining intake and periodic follow-up engagement, the system improves long-term cohort management and enables targeted substudies. The platform also supports integration with clinical studies, as participants are invited through the portal to participate in, for instance, blood bank studies and random control trials, illustrating how CASE allows the digital research infrastructure to connect efficiently with laboratory workflows~\cite{UMCUtrecht2025Biomarkers}.

In a different domain, the Real-Time Response (RTR) mobile application was designed for sentiment analysis during live events with a focus on political debates prior to elections in Germany. It utilizes the survey module from the CASE framework to implement questionnaires and augment them with a real-time feedback data stream, allowing researchers to capture audience reactions dynamically and correlate them with specific debate moments~\cite{Maier2022,TVDuell2025}.

Various applications built on top of the CASE framework, including the listed examples, demonstrate its adaptability across different contexts, from citizen science to mobile app-based real-time sentiment analysis. Although public health has been a central area of application, the underlying architecture is designed to serve a broader research agenda enabling adaptive, scalable data collection systems that prioritize data privacy and offer an accessible and engaging user experience.

\subsection{Validation Through Deployment}
We validate CASE through sustained real-world operation rather than controlled experiments. This approach fits infrastructure research where robustness, maintainability, and institutional fit matter more than isolated performance metrics.

The deployments in Section~\ref{sec:usecases} provide evidence along several dimensions:
\begin{itemize}
    \item Scale and duration: Multiple platforms have operated continuously for several years, collectively engaging tens of thousands of participants, most of them on regular basis (weekly or quarterly).
    \item Adaptability: The same code-base supports influenza surveillance, post-COVID cohort studies, and real-time political sentiment analysis—demonstrating domain-agnostic design.
    \item Institutional adoption: Multiple European countries migrated from legacy Influenzanet infrastructure to CASE, suggesting the framework meets institutional requirements for data sovereignty, regulatory compliance, and long-term maintainability.
    \item Development sustainability: The 2024 architectural rework was motivated by multi-year operation challenges. That we could perform this rework while maintaining backward compatibility demonstrates architectural resilience.

\end{itemize}

Part of validation occurred within EU-funded research projects (Section~\ref{section:projects}) focused on pandemic preparedness. Work within these multinational consortia revealed practical constraints around rapid redeployment, cross-country operation, and long-term sustainability under changing conditions. 
These constraints motivated the architectural changes described in Section~\ref{section:architecture} and development of graphical configuration tools (Section~\ref{section:config-tools}) that enable research teams to create and adapt studies without programming expertise.

In this context, preparedness means more than epidemiological readiness, it refers to the ability of research infrastructures to be rapidly configured, deployed, and adapted, particularly during emerging health threats when external dependencies may be unreliable.

Beyond high-level dimensions, we can trace specific technical capabilities to deployment contexts. Table~\ref{table:evidencemapping} provides this mapping.

\begin{table*}[!ht]
\centering
\caption{Mapping of CASE capabilities to evidence from real-world deployment contexts (Section~\ref{sec:usecases}).}
\label{table:evidencemapping}
\begin{tabular}{p{0.33\linewidth} p{0.63\linewidth}}
\hline
\textbf{Capability} & \textbf{Evidence} \\
\hline
Adaptive study workflows & Used in longitudinal surveillance and cohort follow-up settings . \\

Context-aware survey logic & Individualized follow-ups and state-dependent interactions in production systems. \\

Longitudinal participant state management & Required for sustained cohort participation and multi-study engagement. \\

Integration of temporal and rule-based logic & Used for scheduled interactions and state-dependent follow-ups in deployed systems. \\

Operational maintainability & Architectural redesign motivated by multi-year operation across multiple platforms. \\
\hline
\end{tabular}
\end{table*}

This validation approach does not provide controlled comparisons with alternative platforms, formal usability metrics, or quantified measures of configuration effort. Future work should include systematic usability studies and reproducible benchmarks for common participatory surveillance workflows.

\subsection{Lessons Learned}
The development and rework of the CASE platform and its deployment in multiple countries and institutions provided practical insights. The following selected lessons learned offer guidance for future implementations of participatory data collection platforms, particularly in preparedness contexts.

\paragraph{Customization Capability as Core Value}
Our experience working closely with stakeholders demonstrated that the primary benefit of using our study framework compared to other alternatives, which in some cases might be easier to deploy, was the ability to \emph{"make things happen"}. When using off-the-shelf software or services (e.g., as SaaS products), changes to the workings, design, and functionality are very limited if possible at all. While our framework allows a huge number of customizations through the standard feature set, it also offers extendability to build around the core or compose new components entirely.

\paragraph{Reducing Technical Barriers Through Tooling}
The developed graphical configuration tools address a critical preparedness challenge: enabling rapid study deployment by teams without programming expertise. Survey creation previously required manual coding of survey definitions, but researchers can now build surveys through the visual Survey Editor with structured item editing and real-time preview. Similarly, study logic and display conditions can be created with the Expression Editor instead of manual scripting, speeding up iteration during time-sensitive deployments. While these tools significantly improve accessibility, infrastructure deployment still requires IT support, highlighting the ongoing tension between usability and technical complexity in participatory research platforms.

\paragraph{Preparedness Requires Balanced Standardization and Flexibility}
Preparedness, which seems to grow in importance, requires both standardization for rapid deployment and flexibility to adapt to evolving situations. New studies typically want to accommodate specific scenarios as quickly as possible. A certain degree of standardization is necessary for the quick deployment and established processes, but a high degree of flexibility is also required to react to the dynamic nature of evolving situations (such as in a pandemic). In addition, stakeholders often have conflicting preparation priorities, ranging from technical infrastructure to study design readiness to ethical compliance.

\paragraph{Digital Sovereignty as Strategic Necessity}
Additionally, compared to many services offered, digital sovereignty is a key aspect. Having the ability to host, maintain, or develop the solution if needed without external entities. Although our team currently provides efficient engineering solutions, the open source and permissive nature of the project ensures the independence of the organizations. This autonomy becomes particularly critical during crisis situations, when external dependencies may become unreliable or unavailable.

\paragraph{Organizational and Training Requirements}
Successful deployment requires more than technical solutions. Even user-friendly platforms need structured onboarding and training programs. We found that iterative development cycles that involved technical teams, researchers, and ethics/privacy officers were essential for addressing real-world constraints.
A deep understanding of the actual research goals proved critical. Effective solutions emerged only when developers truly grasped the surveillance objectives and priorities. This required sustained collaboration with domain experts throughout development, not just during requirement gathering. Without this ongoing dialogue, technical teams risk building features that seem logical but miss actual research needs.

\paragraph{Sustainability and Funding Challenges} 
Software maintenance is an ongoing necessity as dependencies, libraries, and security requirements evolve constantly. However, this creates a fundamental tension as often research projects need working solutions immediately but rarely budget for long-term maintenance. Funding typically supports specific research goals, not the "invisible" work of keeping infrastructure up to date and secure, which threatens sustainability. Although organizations benefit from the framework today, few contribute to its future viability. Ensuring long-term availability requires new funding models that recognize infrastructure maintenance as essential research support, not optional overhead. Without sustainable funding for core development and governance, even successful platforms risk obsolescence.

\section{Closing Remarks and Next Steps}
We presented CASE, an open-source framework for participatory adaptive research built on over a decade of disease surveillance experience. Designed to meet the demands of public health infrastructure, CASE has demonstrated robustness in large-scale real-world deployments, including national COVID-19 monitoring platforms and longitudinal post-infection cohort studies. Additionally, CASE is well suited for general-purpose use. Its modular event-driven logic, privacy-aware design, and highly customizable workflows make it suitable for behavioral science, environmental health, and other domains where context-aware participatory data collection is critical. 

The 2024 rework ensures that CASE remains aligned with modern standards and research needs, significantly enhancing both maintainability and deployment flexibility. The new architecture retains flexibility and scalability, ensuring that the system remains relevant and extendable for future applications. By combining technical robustness with a user-friendly experience for participants and ethical design, CASE establishes itself as a research infrastructure capable of supporting both urgent epidemiological surveillance and long-term interdisciplinary studies. The recent adoption of CASE by numerous European platforms, particularly within the \emph{Influenzanet} network, highlights its growing role as a foundational infrastructure for multi-study participatory research.

Beyond its technical capabilities, CASE addresses the critical need for institutional control in research infrastructure.  Its open-source and self-hostable nature allows organizations to retain sov\-ereignty over their data collection systems when needed, without mandatory dependence on external commercial platforms or services.

Several limitations remain. We lack formal usability evaluations of participant and researcher experience across deployments. The monolithic redesign improved maintainability, and we developed graphical configuration tools (Survey Editor, Expression Editor, Admin UI) that enable researchers to create studies without coding. However, initial deployment and infrastructure setup still require IT expertise, limiting accessibility for some institutions. Most critically, long-term sustainability depends on continued institutional support and formal governance, not just technical design. Without sustainable funding for documentation, maintenance, and institutional handover, even successful platforms risk obsolescence between crises.

To address these challenges, we will consolidate and stabilize existing tooling rather than adding new features. In practice, this means improving documentation, simplifying configuration workflows, and further refining the user-facing tools that already support study setup and deployment. Our aim is not to remove flexibility, but to make common workflows easier to understand, reproduce, and adapt by teams without extensive technical expertise.

Based on repeating patterns observed across deployments, we also plan to provide a small set of example applications that can serve as starting points for common participatory research scenarios, such as longitudinal cohort studies or weekly surveillance reports. Rather than acting as fixed templates, these examples are intended to capture design decisions and best practices that have proven useful in practice. In parallel, we are exploring whether a hosted deployment model could be useful for some organizations without requiring technical expertise or proper in-house infrastructure, while still preserving CASE's emphasis on self-hostability and data sovereignty.

\bibliography{references}

\end{document}